\documentclass[singlecolumn,letterpaper,12pt]{article}
\pdfoutput=1
\usepackage{graphicx}
\usepackage{bm}
\usepackage{graphicx}
\usepackage{latexsym}
\usepackage{amsmath}
\usepackage{amsthm} 
\usepackage{times}

\usepackage{txfonts}
\usepackage[bookmarks=false]{hyperref}
\usepackage{stfloats}
\usepackage{algorithm}
\usepackage{algorithmic}

\begin{document}

\date{}

\title{VM Power Prediction in Distributed Systems for Maximizing Renewable Energy Usage}
\author{
Ankur Sahai \\
University of Mainz, Germany
}

\hyphenpenalty=1000
\setlength{\parindent}{10pt}
\setlength{\parskip}{1ex} 
\maketitle
\thispagestyle{empty}
\noindent
\section{Abstract}
\label{sec:abstract}

In the context of GreenPAD project it is important to predict the energy consumption of individual VMs / workload for optimal scheduling (running those VMs which require higher energy when there is more green energy available and vice-versa) in order to maximize green energy utilization.

This involves the following tasks:

\begin{enumerate}
\item VM energy measurement for different configurations (flavor + workload)
\item VM energy prediction for a new configuration.
\end{enumerate}
We propose a model for predicting the power usage of the VMs based on regression. We first collect the resource usage and the associated power usage for different VM configurations and use this to build a regression model. Then we use the information about the resource usage patterns of the new workload to predict the power usage.

\section{Setup}
\label{sec:setup}

For modeling a particular workload / VM we run it on an isolated server and measure the resource and power usage (after recording the static power usage without VM and finding the difference). This is achieved by the means of a bash script which launches a VM instance (created from a VM image which has workload simulation tool - Stress preinstalled) with a specific flavor type on a particular host using OpenStack Nova commands, executes a workload on the VM and measures the resource and associated power. The resource usage information is collected using \emph{collectd} daemon running on the host. Power consumption is measured using the IPMI tool via the \emph{IPMItool}. We verify the power usage using a physical power meter - \emph{Gude} which is connected to a network tool called Spider which can be access using a JNLP based front end on a web-browser.

The bash script first launches a VM on a host and starts the collectd service on the host. The VM is assigned an external IP using a nova command. Then a workload is started using ssh on the remote VM. Simultaneously the IPMI tool is also started in the background which keeps recording the power usage on the host inside a while loop running as a separate thread in parallel. After the experiment terminates the collectd service is stopped on the host and the thread for power collection is terminated. Later on we use this data - resource usage and power to build a regression model. As the data collected by the collectd is written in a RRD (Round Robin Database) we use a tool called - \emph{rrdtool} to read values from the \emph{rrd} database and plotting graphs. Start and end time of the experiments is also noted for each experiment. Sleep commands are used inside the bash scripts e.g. after creating a VM.

Automation bash script calls the Nova APIs for creating, deleting, associating an external IP etc. SSH command is used to execute the workload on the VM which is preinstalled with the tool. System time is extracted upto second precision at the start and end of the experiment. In order to execute the Nova commands first a source command is used to publish the \emph{novarc} and \emph{openrc-admin} files. \& operator is used to launch a separate thread before launching the experiment, for collecting the power data using IPMI command. Linux top command is used to monitor the KVM process and its associated resource usage. RRDtool graph command is used to print graphs from the RRD files written by collectd. After the experiment is over, the thread for power collection is terminated using the Linux kill command and the data from the file is parsed using the Matlab text read command. Plot command is used to generate the plot from the power data file. The file containing the start and end time of the experiment is also parsed to add the appropriate labels to the power plot in order to make it comparable to the resource usage plots.

In the collectd configuration file the interval for data collection is set to 1 second. Apart from this the CPU, Memory, Disk and Network plugins are enabled using LoadPlugin command. Within the Network plugin the specific interface which has to be monitored is also specified.While parsing the RRD files using the RRDtool the value of the sensor such as Value(for CPU and Memory), Receive and Transmit (for Network) and Read and Write (for Disk). Similarly for the RRDtool graph command the name of the sensor apart from the color of the plot and the type - line / area and a label / legend can be specified.

The main aim of the script is to launch VMs with different configuration which includes flavor type 

Collectd tool can be configured to collect different metrics such as CPU user, idle and wait time, Memory - used, free, Disk - byte rate for read and write and Network - byte rate for receive and transmit operations. Collectd comes with a front end - CGI based tool called \emph{collection3}. 

VM resource consumption is also verified by monitoring the KVM process associated with a VM. Using the Linux top command it is possible to retrieve the percentage resource usage as well as the total resource available and this information is used to derive the actual resource usage.

The default VM image is the Ubuntu based Maverick image which has the stress tool preinstalled on it (by creating a snapshot after installing the stress tool). The underlying hardware consists of Fujitsu servers running \emph{Openstack} cloud platform and the default hypervisor is \emph{Qemu} / \emph{KVM} based. The server hardware consists of four processors of the type Intel(R) Xeon(TM) CPU 3.00GHz. The hard disk has a capacity of 250 GB with Ubuntu server edition running on top. \cite{vm-power-meter} considers each VM as a process and the processes running inside as threads.

In addition to the stress tool we use custom bash scripts. For generating a CPU intensive workload we consider a bash script with infinite while loop which executes addition of two numbers. For memory intensive workload we consider the recursive fork operation using a bash script and the \emph{\&} operator which launches multiple threads. Recursive fork operation causes an exponential increase in memory consumption as each newly created thread is assigned an address space of its own. For I/O intensive operation we use the \emph{dd} command in Linux that reads data from \emph{/dev/random} (default device file used to generate random data) and writes to \emph{/dev/null} (default device file used to provide a character stream for storing data). For network intensive workloads we send / receive files using FTP protocol between VMs as we think it is a simple yet efficient test for networking.

The collected power and VM configuration data is also stored in a MySQL database. This data stored in the MySQL database is queried for predicting the VM energy consumption later on. In the end we query the database for the resource (CPU, Memory, Disk and Network) and corresponding power usage to build a regression model using Matlab.

Apart from this the energy consumption was also measure while running multiple VM templates without any workload. No significant difference in power was observed while launching upto two VMs.

A sample call to stress looks as follows: stress --cpu 8 --io 4 --vm 2 --vm-bytes 128M --timeout 10s where 8, 4, 2 are the number of threads for CPU, I/O and memory and 128M is the size of the memory block that is be used and freed whereas 10s is the time of the run. The CPU thread spins on a SQRT operation of a random number. I/O thread calls the sync() function periodically (spinning) which writes the data buffered in memory to the disk. Memory thread frees and allocates memory in cycles.

Some of the other tools used were: CPUStress, DBench, Bonnie, Netperf, UnixBench, WebStone (which is a web application simulation tool).

\begin{figure}
\centering
\includegraphics[width=150mm]{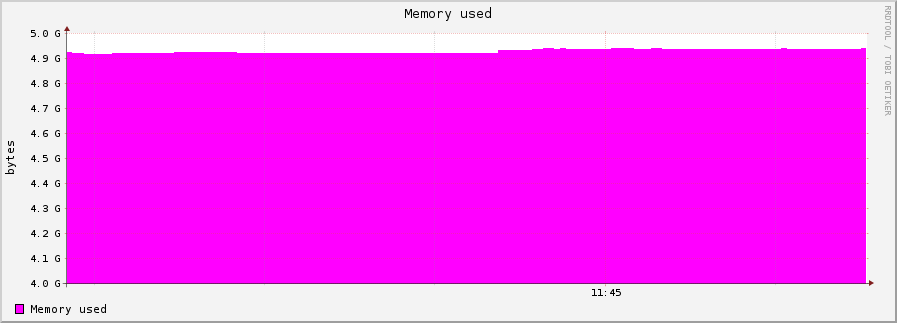}
\caption{Memory used in bytes}
\label{memory-used}
\end{figure}

\begin{figure}
\centering
\includegraphics[width=150mm]{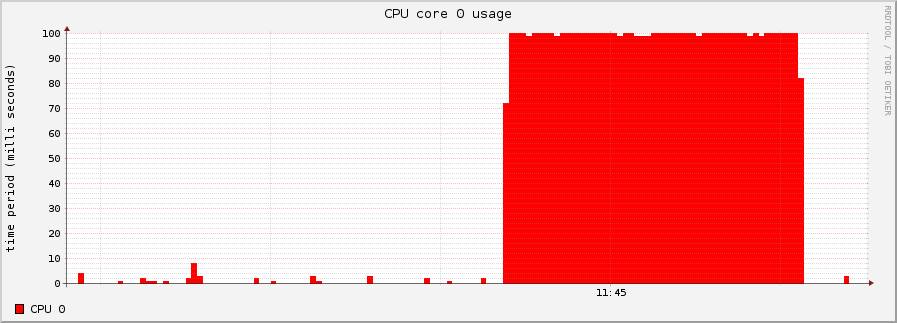}
\caption{CPU usage time period in milliseconds}
\label{cpu-time-period}
\end{figure}

\begin{figure}
\centering
\includegraphics[width=150mm]{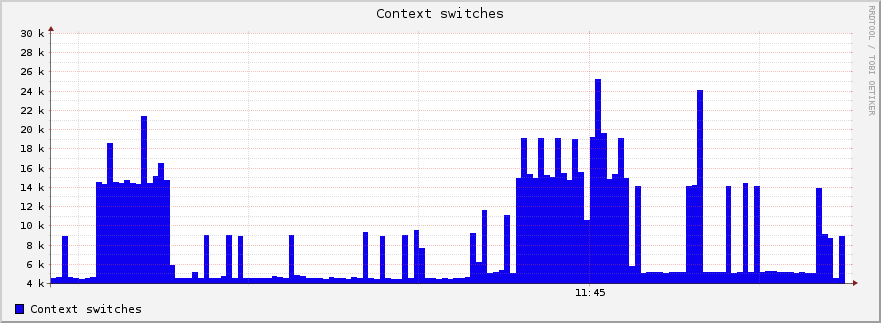}
\caption{Number of context switches}
\label{context-switches}
\end{figure}

\begin{figure}
\centering
\includegraphics[width=150mm]{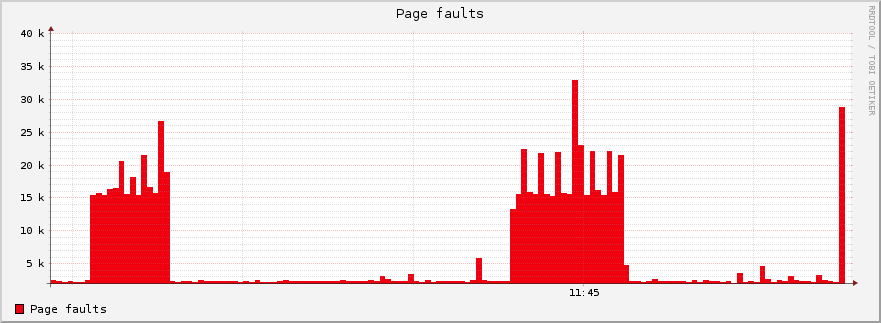}
\caption{Number of page faults}
\label{page-faults}
\end{figure}

\begin{figure}
\centering
\includegraphics[width=100mm]{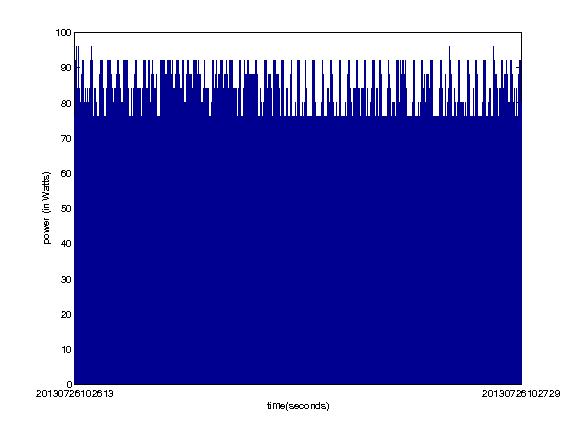}
\caption{Power usage measured using IPMI}
\label{power}
\end{figure}

Some of the sample output received using collectd tool is attached hereunder for the experiment running around the label 11:45 on the X-axis of the graph. Spikes in the network bytes transmitted are due to the SSH command that is used to launch the workload on the VM. Memory used Figure \ref{memory-used} shows as almost constant because there are multiple threads that are doing malloc and free operations. Memory caches remains almost constant for the same reason as in \cite{memory-used}. Bytes written to disk shows spikes as the sync() operation writes the data collected to the disk which corresponds to the malloc() operation. The continuos increase in the CPU usage values in Figure \ref{cpu-time-period} is due to the the behavior of the thread that continuously loops on the SQRT() operation. The number of context switches Figure \ref{context-switches} and page faults Figure \ref{page-faults} peak at the start of experiment and are almost null during the later stages of the experiment.  A sample power plot is shown in Figure \ref{power}. Power which is measured in Watts using IPMI tool increases by steps of 4 W. The periods of higher power correspond to the time when experiment was running.

\begin{figure}
\centering
\includegraphics[width=150mm]{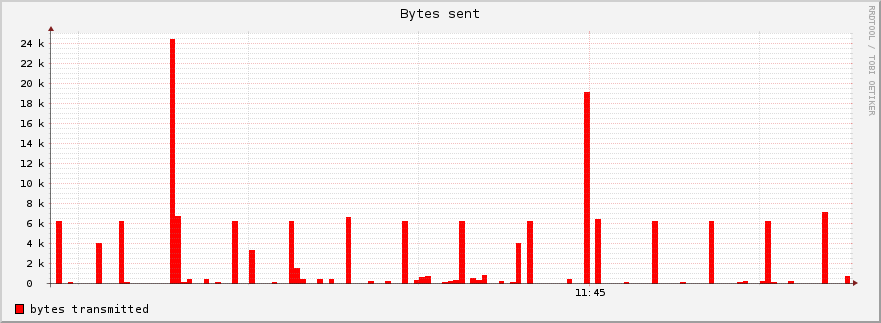}
\caption{Bytes transmitted by the network interface}
\label{network-bytes-transmit}
\end{figure}

\begin{figure}
\centering
\includegraphics[width=150mm]{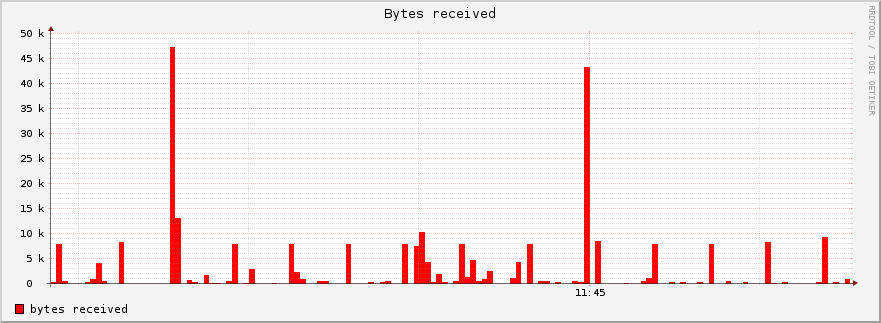}
\caption{Bytes received by the network interface}
\label{network-bytes-receive}
\end{figure}

\begin{figure}
\centering
\includegraphics[width=150mm]{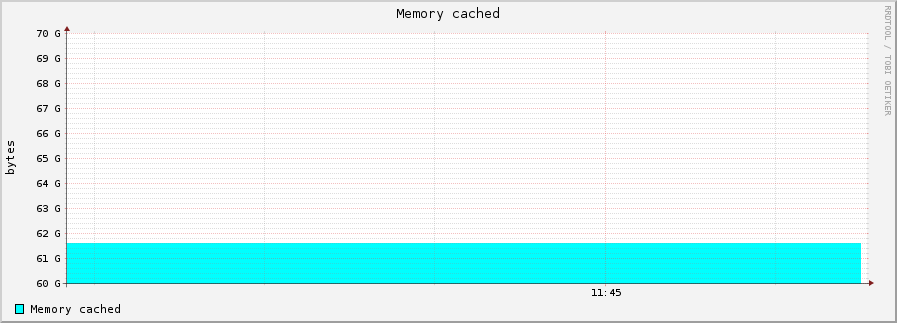}
\caption{Memory cached in bytes}
\label{memory-cached}
\end{figure}

\begin{figure}
\centering
\includegraphics[width=150mm]{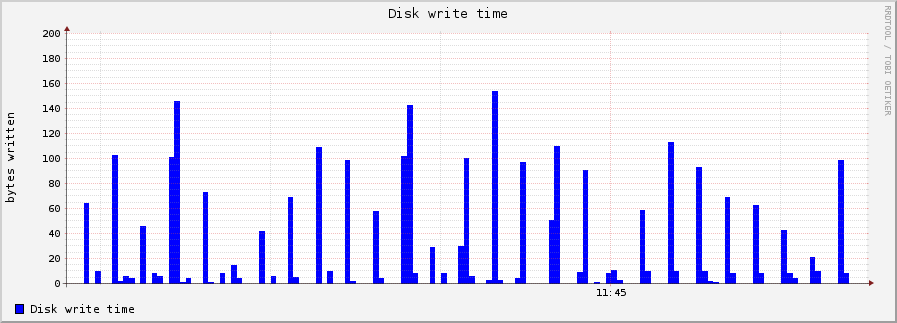}
\caption{Bytes written to disk}
\label{disk-bytes-written}
\end{figure}

\section{Machine learning tools}
\label{sec:ML-tools}
Some of the ML libraries tried were: Weka and libSVM. An implementation of Weka called SysWeka can be found here. Some of the routines implemented in the Bayes classifiers are Naive Bayes and others are regression based models for workload classification.

\section{Heterogenous Workload mixture}
\label{sec:hetro-workload-mix}
We also analyze the power consumption for different type and number of workloads. The different types of workloads are simulated using bash scripts (for specific workload characteristics). VM instances selected were of type m1.small (8GB memory, 10GB storage, 4 vCPUs). The challenge in measuring the total power is because the average individual powers of the different workloads may not add up because of the resource sharing and access behavior amongst multiple VMs which may often times produce a total consumption less than the total. As the power cap of the server was reached by running the tests mentioned in Section \ref{sec:setup} it was not possible to accurately measure the total power consumption. We are currently in the process of identifying the suitable tests for measuring the power consumption for different mix of workloads.

\section{Results}
\label{sec:results}
Based on the experiments described in Section \ref{sec:setup}, we notice that CPU intensive operations have the highest power consumption. This is closely followed by a memory operation. Network intensive operation have the next highest power consumption rate (for higher values of maximum bit transfer rate $>$ 100 KB/sec). The least power usage was for the disk I/O operations. We observe that the CPU, Memory, Storage and Network intensive operations consume approximately 30 - 40 W, 24-28 W, 8-10 W and 22 -32 W respectively. These values are close to the results in \cite{vm-power-meter} which use a similar platform with KVM / Qemu based hypervisor.Power for the peak load was around 90-100W measured using the \emph{stress} tool.

We believe that the peak power consumption for the VMs running on a cloud can be used as the coarse-grained scheduling strategy and the power prediction using the bayesian model can be used for finer level scheduling.


\begin{thebibliography}{99}
\bibitem{joulemeter}
Aman Kansal, Feng Zhao, Jie Liu, Nupur Kothari, and Arka A. Bhattacharya. 2010. Virtual machine power metering and provisioning. In Proceedings of the 1st ACM symposium on Cloud computing (SoCC '10). ACM, New York, NY, USA, 39-50. DOI=10.1145/1807128.1807136 http://doi.acm.org/10.1145/1807128.1807136

\bibitem{hypervisor-vm-resource-track}
Stoess, Jan, Christian Lang, and Frank Bellosa. "Energy management for hypervisor-based virtual machines." Proceedings of the USENIX Annual Technical Conference. 2007.

\bibitem{vm-power-meter}
Bhavani Krishnan, Hrishikesh Amur, Ada Gavrilovska, and Karsten Schwan. 2011. VM power metering: feasibility and challenges. SIGMETRICS Perform. Eval. Rev. 38, 3 (January 2011), 56-60. DOI=10.1145/1925019.1925031 http://doi.acm.org/10.1145/1925019.1925031

\bibitem{multiple-workload-energy}
Jeonghwan Choi; Govindan, S.; Urgaonkar, B.; Sivasubramaniam, A.; , "Profiling, Prediction, and Capping of Power Consumption in Consolidated Environments," Modeling, Analysis and Simulation of Computers and Telecommunication Systems, 2008. MASCOTS 2008. IEEE International Symposium on , vol., no., pp.1-10, 8-10 Sept. 2008
doi: 10.1109/MASCOT.2008.4770558

\end{thebibliography}
\end{document}